# Geert Hofstede et al's Set of National Cultural Dimensions - Popularity and Criticisms

**Kiril Dimitrov**[*]

## Summary

This article outlines different stages in development of the national culture model, created by Geert Hofstede and his affiliates. This paper reveals and synthesizes the contemporary review of the application spheres of this framework. Numerous applications of the dimensions set are used as a source of identifying significant critiques, concerning different aspects in model's operation. These critiques are classified and their underlying reasons are also outlined by means of a fishbone diagram.

## Introduction

For more than thirty years a set of cultural dimensions, proposed by Hofstede et al and constructed as a result of their continuous research in identifying and explaining cultural differences at the national and regional level, has attracted the attention of different social actors – scientists, managers, politicians, administrators, opinion leaders, and other agents, because potential cultural differences are observed to have influenced dominating organizational practices and theories in the context of increasing globalization and economic turbulence. Furthermore, the contemporary times may be characterized by realization of intensive interactions between differing cultures, "traversing national borders, co-mingling, hybridizing, morphing, and clashing through media, migration, telecommunications, international trade, information technology, supranational organizations, and unfortunately terrorism" (Nakata, 2009, p.4) which situation serves as a catalyst for the unceasing interest in Hofstede's research results.

Naturally, this lasting memory of the aforementioned cultural dimensions set is deeply grounded in the times of its creation, because the Dutch scientist even in the early 1980s proposed a plausible explanation for the great significance of the "nationality – management" relationship, formulating three reasons (Hofstede, 1983):

- The political reason is justified by essence and basic characteristics (for example institutions, ways of using them) of the 'nation' construct.
- The sociological reason relates to the special way of how people perceive and what value they ascribe to their identity and sense of belonging, which certainly directs their behaviors in key situations and may possibly cause the demonstrated extremes in their decision-making (for instance to go to war).

[*] Associate professor, Ph.D., "Industrial business" department at the University of National and World Economy, e-mail: kscience@unwe.eu

- The psychological reason is used to reveal the influence of national culture factors on human thinking, expressed by one's specific childhood and adolescence learning experiences in diverse cultural milieus as separate families, schools and organizations.

A richer and contemporary 'official justifications' of the observed popularity for this model are grounded by Hofstede and Minkov (2011), but these will not be dwelled on here because 'user experience' is considered of greater importance in this deliverable. The unceasing interest in applying and appraising Hofstede's cultural dimensions set by different constituencies and the contemporary business environment conditions (uncertainty, instability, unpredictability, ambiguity, etc.) in which the organizations are operating today are the two main factors that provoked my scientific interest in making a historical review and taking an up-to-date snapshot of this cultural model in an attempt to: (a) reveal important nuances in its structural development, (b) trace the accumulation of its application spheres, and (c) analyze the criticisms related with it. The current article represents the means of achieving the aforementioned goals.

## The set of national culture dimensions as a moving target

The cultural model proved to be a moving target in the analyzed period (since 1980 up to now) in terms of at least two perspectives – its structure and main application spheres. The first perspective of the model development seems to be a dependent variable on Hofstede et al's investigative questions and subsequent research actions. The second perspective depends on the successive activities of other scientists and consultants who tried, and are still trying to apply this model to specific fields of management and other social sciences.

Initially Hofstede's research results on national and regional cultural differences emerged as a set of four dimensions. Later on by extending his research and collaborating with other scientists the Dutchman gradually enriched his model to six dimensions that is evidenced in a number of his publications (see table 1). The contributions of the Canadian psychologist Michael Harris Bond and Michael Minkov – a Bulgarian researcher in the fields of ancient languages, anthropology, and management sciences, may be considered as key marker events in the model's elaboration (Bergiel, Bergiel, Upson, 2012; Hofstede, 2011; Adolphus, 2011; Hofstede, Hofstede, 2014; Hofstede, Hofstede, Minkov, 2010), as follows:

- In turn Michael Harris Bond succeeded in adding a new element to the model's structure – the so called fifth dimension, first labeled as Confucian Dynamism, but later on refined as Long versus Short Term Orientation. It is the result of a comprehensive study of Chinese Values, conducted in the Asian-Pacific region (23 countries, research units: students in psychology, fifty men and fifty women in each country). The replications of three dimensions were observed between IBM research and the Chinese Value survey. The latter did not provide any evidence of "uncertainty avoidance" in this region. Yet at that time the fifth cultural dimension was accepted as a Chinese artifact.
- Michael Minkov's scheme to contribute: The scholar used data provided by the World Values Survey, the United Nations

*Table 1. Mapping the structure of Hofstede's cultural dimensions model*

| Structural versions | Cultural dimensions with short descriptions* | Respective references |
|---|---|---|
| Initial structure, containing four dimensions | 1. *Individualism versus Collectivism (IC)*. The two constructs are contrasted by incarnated respective meanings. Individualism refers to the extent to which people are expected to stand up for themselves and to choose their own affiliations. Such societies are characterized by loose ties between individuals, expressed in the basic need of everyone's taking care of her/himself and her/his immediate family. Collectivism refers to the extent to which people are expected to act predominantly as members of a life-long group or organization. Such societies are characterized by people's integration since birth into strong, cohesive in-groups (for example extended families) which continue protecting them in exchange for unquestioning loyalty. There is no political shade in the meaning of "collectivism". The difference between the two constructs may be measured as the degree to which individuals are integrated into groups.

2. *Large or Small Power Distance (PD)*. The degree to which a society accepts and expects there to be differences in the levels of power in organizations and institutions (for example the family). A high score suggests that there is an expectation and acceptance that some individuals wield larger amounts of power than others. A low score reflects the view that all people should have equal rights. The inequality is defined from below, not from above. Followers and leaders may appraise a society's level of inequality. The comparative analysis between different societies is important, concerning power and inequality, because the analysis relies on the axiom that "all societies are unequal, but some are more unequal than others".

3. *Strong or Weak Uncertainty Avoidance (UA)*. The extent to which a society accepts uncertainty, ambiguity and risk, i.e. its members attempt to cope with anxiety by minimizing them or to what extent they feel either uncomfortable or comfortable in unstructured situations that may be described as novel, unknown, surprising and different from usual. Uncertainty avoiding cultures try to minimize the possibility of such situations by introducing and maintaining strict laws and rules (i.e. there is only one truth), safety and security measures. The members of such cultures are more emotional, motivated by inner nervous energy, and tend to remain longer with their present employer. Uncertainty accepting cultures are characterized by greater tolerance of different opinions/religions/ philosophies and establishment of few rules as possible, greater phlegmaticness, contemplativeness and keeping in control of human emotions.

4. *Masculinity versus Femininity (MF)*. These are two extremes, depicting distribution of emotional roles between the genders. Male values (the assertive pole) include(s) competitiveness, assertiveness, ambition, and the accumulation of wealth and material possessions, whereas feminine values (the modest pole) are (is) oriented to relationships and quality of life. Furthermore, several research results are important here: (a) women's values differ less among societies than men's values; (b) men's values from one country to another may vary from very assertive and competitive and maximally different from women's values on the one side, to modest and caring and similar to women's values on the other. (c) women in feminine countries have the same modest, caring values as the men while in masculine countries they are more assertive and more competitive, but not as much as the men. | 1. Hofstede, G., Culture's Consequences: International Differences in Work-Related Values. Beverly Hills/ London: SAGE Publications, 1980.

2. Hofstede, G., "Dimensions of National Cultures in Fifty Countries and Three Regions." In Expiscations in Cross-Cultural Psychology, edited by J. Deregowski, S. Dziurawiec, and R. C. Annis. Lisse, Netherlands: Swets and Zeitlinger, 1983. |

*Table 1. Mapping the structure of Hofstede's cultural dimensions model (continued)*

| Structural versions | Cultural dimensions with short descriptions* | Respective references |
|---|---|---|
| Newer structure, containing five dimensions | 1. Individualism versus Collectivism.<br>2. Large or Small Power Distance.<br>3. Strong or Weak Uncertainty Avoidance.<br>4. Masculinity versus Femininity.<br>(already described)<br><br>5. *Long versus short term orientation (TO)*. It deals with a society's „time horizon" or the importance attached to the future versus the past and present. In long term oriented societies pragmatic virtues oriented to future rewards as thrift, perseverance and adaptation to changing circumstances are valued more. In short term oriented societies virtues, related to respect for tradition, national pride, preservation of „face", social obligations, reciprocation of gifts and favors are valued more. | 1. Hofstede, G. and M. H. Bond (1984). "Hofstede's Culture Dimensions: An Independent Validation Using Rokeach's Value Survey." Journal of Cross-Cultural Psychology 15(4): 417-433.<br>2. Hofstede, G. and M. H. Bond (1988). "The Confucius connection: From cultural roots to economic growth." Organizational Dynamics 16(4): 5-21.<br>3. Hofstede, G. (1991). Cultures and Organizations: Software of the mind. London, McGraw-Hill.<br>4. Minkov, M. (2007). What makes us different and similar: A new interpretation of the World Values Survey and other cross-cultural data. Sofia, Bulgaria: Klasika i Stil. |
| The newest structure, containing six dimensions | 1. Individualism versus Collectivism.<br>2. Large or Small Power Distance.<br>3. Strong or Weak Uncertainty Avoidance.<br>4. Masculinity versus Femininity.<br>5. Long versus short term orientation.<br>(already described)<br><br>6. *Indulgence versus Restraint (IR)*. Indulgence is typical of a society, allowing relatively free gratification of basic and natural human drives related to enjoying life and having fun. Restraint stands for a society, suppressing gratification of needs and regulating it by means of strict social norms. | 1. Hofstede, G., Hofstede, G. J. & Minkov, M. (2010). Cultures and Organizations: Software of the Mind (Rev. 3rd ed.). New York: McGraw-Hill.<br>2. Minkov, M. (2007). What makes us different and similar: A new interpretation of the World Values Survey and other cross-cultural data. Sofia, Bulgaria: Klasika i Stil. |
| The presented descriptions are based on the following sources: (Hofstede, Hofstede, 2014; Hofstede, Hofstede, Minkov, 2010). | | |

Organization and the World Health Organization in order to construct four dimensions - industry versus indulgence, monumentalism versus flexumility, hypometropia versus prudence, and exclusionism versus universalism. Later on he adapted his research findings to enrich in an appropriate way (statistically and conceptually) Hofstede's model of cultural dimensions on national level. Thus "indulgence versus restraint" dimension came into being which the author considers similar to the earlier proposed "industry versus indulgence". Furthermore, the researcher confirmed the utility and universality of the fifth dimension – "long versus short term orientation" in the Asian-Pacific region by discovering a useful analogue in the World Values Survey.

This is how the contemporary structure of cultural dimension set gradually took its current shape (see table 1). Furthermore, the observed widespread adoption of Hofstede's dimensions sounds even better explained through the standpoint of "an intelligent user" (Chanchani, Theivanathampillai, 2009), as follows: (a) design of a clear framework, intended to classify diverse cultures, due to deliberate integration of previously fragmented cultural constructs and theories; (b) perceived simplicity in the application of these cultural dimensions by users from business world and academics; (c) a new value measurement technique is brought to our attention, which is not a frequent phenomenon; (d) meeting researchers'demands by offering an extensive data set for empirical analysis.

The second perspective in the elaboration of Hofstede's cultural dimensions set may be outlined by tracing its possible application spheres through key specific studies and summarizing studies, intended to provide reviews of publications from different periodicals and/or different scientific databases for certain time periods. Most frequently Hofstede's cultural model simplicity to use and the ease of comparability, allowed by the utilization of a quantitative measure of culture, are pointed as basic reasons for its great popularity and high utility among academics and in the business field (see Bing, 2004; Hoppe, 2004; Sivakumar, Nakata, 2001). This is the reason why Kirkman, Lowe and Gibson (2006) succeed in their endeavors to identify shades of use for Hofstede's model in researches, conducted by other scientists, and classify them by two criteria. In fact the scientific team reviewed 180 studies, published in 40 business and psychology journals and two international annual volumes between 1980 and June 2002. The structure of their classification system seems a bit complex, because it is designed in two tiers (see table 2):

- The first criterion appraises the role of cultural values in investigated relationships. The two poles along the chosen

*Table 2. Kirkman, Lowe and Gibson's classification scheme of literature review with the respective number of included articles*

| | Individual level | Gnjup/organizatiori level | Country level | Total |
|---|---|---|---|---|
| Culture as a main effect | 64 | 6 | 78 | 148 |
| Culture as a moderator | 23 | 5 | 4 | 32 |
| Total | 87 | 11 | 82 | 180 |
| Note: If a study was listed in more one section, it was counted only once in the section in which it appeared. | | | | |

*Source: Kirkman, Lowe and Gibson (2006)*

*Table 3. Kirkman, Lowe and Gibson's grid of research subject matter by level of analysis)*

| Level of analysis / Management and applied psvcliology domains | Individual | | Group/ organization | | Country | |
|---|---|---|---|---|---|---|
| | main[4] | mod** | main | mod | main | mod |
| *Change management* | 4 | 0 | 0 | 0 | 0 | 0 |
| *Conflict management* | 4 | 0 | 2 | 0 | 1 | 0 |
| *Decision-making* | 4 | 0 | 0 | 0 | 0 | 0 |
| *Human resource management* | 5 | 0 | 0 | 0 | 4 | 0 |
| *Leadership* | 4 | 1 | 1 | 0 | 3 | 0 |
| *Organizational citizenship behavior (OCB)* | 2 | 0 | 0 | 0 | 0 | 0 |
| *Work-related attitudes* | S | 6 | 0 | 0 | 7 | 2 |
| *Negotiation* | 9 | 2 | 0 | 0 | 0 | 0 |
| *Reward allocation* | 8 | 0 | 0 | 0 | 0 | 0 |
| *Behavior relating to group processes and personality* | 16 | 4 | 3 | 6 | 2 | 0 |
| *Entrepreneursh ip* | 0 | 0 | 2 | 0 | 1 | 0 |
| *Social networks* | 0 | 0 | 0 | 0 | 2 | 0 |
| *Entiy modes* | 0 | 0 | 0 | 0 | 21 | 1 |
| *Foreign direct investment* | 0 | 0 | 0 | 0 | 6 | 0 |
| *Joint venture characteristics and performance* | 0 | 0 | 0 | 0 | 18 | 1 |
| *Alliance formation* | 0 | 1 | 0 | 0 | 2 | 1 |
| *Innovation and research and development* | 0 | 0 | 0 | 0 | 4 | 0 |
| *Societal outcomes (e.g., wealth, national accounting systems, number of intellectual property violations* | 0 | 0 | 0 | 0 | | 2 |
| *Motivation* | 0 | 5 | 0 | 0 | 0 | 0 |
| Organizationa 1 justice | 0 | 5 | 0 | 0 | 0 | 0 |
| Adapted from: Kirkman, Lowe and Gibson (2006). Legend: *MAIN - main effect study; **MOD - moderating effect study. | | | | | | |

continuum, used to identify two groups of studies, are occupied by: (a) examination of main associations between values and outcomes (main effect studies, labeled as 'Type I'), and (b) revelation of cultural values as moderators (moderator studies 'Type II'). This criterion is borrowed from a research, conducted by Lytle et al. (1995).

- The second criterion is formulated on the preferred level of analysis in the reviewed studies, i.e. individual, group/ organizational, or country. This choice is deliberate, because on one side, the authors are aware of Hofstede's limiting the application for his framework only to country and regional level studies while on the other side, they consider the availability of many studies where cultural dimensions are adapted for implementation on individual or group/organization levels. Such 'broadening of the research horizons' may be justified through posing of appropriate research questions and the extent of greater commonality within surveyed groups, than between them (Sivakumar, Nakata, 2001). Thus the authors create a grid of research themes that have been attracting the attention of their colleagues as application spheres for Hofstede's culture dimensions set, layered by the preferred levels of analysis (see table 3).

The majority of researchers who kept

Hofstede's recommendations about the appropriate level of analysis demonstrate greater interest in application spheres for cultural dimensions as "entry modes", "joint venture characteristics and performance", "societal outcomes", and "work-related attitudes". The group of scientists who sought new applications of Hofstede's model on individual level showed keen interest in "behavior relating to group processes and personality", "negotiation", "reward allocation" and "work-related attitudes". That is why it is not surprising that certain changes into Hofstede's framework are proposed, so it may be applied without allowing "ecological fallacy" on the individual level of analysis (see Grenness, 2012). As a whole the approach of presenting culture as a main effect dominates on all identified levels of analysis and within most of the target research domains with the exception of "behavior relating to group processes and personality" on group/ organization level, as well as "motivation", "organizational justice" and "alliance formation" on individual level, within which issues were investigated by using culture as a moderator. Furthermore, Kirkman, Lowe and Gibson's (2006) survey results reveal the stronger interest, demonstrated by scientists to cultural dimensions as "individualism-collectivism" and "power distance" (see table 4). It should be noted that logically this research does

Table 4. Kirkman, Lowe and Gibson's count of cultural values inclusions by type of effect and level of analysis

|  | Individualism–collectivism | Power distance | Uncertainty avoidance | Masculinity–femininity | Confucian dynamism |
|---|---|---|---|---|---|
| Main: individual | 58 | 11 | 8 | 8 | 3 |
| Main: group/organizational | 8 | 1 | 1 | 1 | 0 |
| Main: country | 27 | 27 | 26 | 20 | 2 |
| Moderating: individual | 19 | 9 | 3 | 3 | 0 |
| Moderating: group/organizational | 5 | 1 | 0 | 0 | 0 |
| Moderating: country | 3 | 2 | 1 | 1 | 0 |

Source: Kirkman, Lowe and Gibson (2006)

not include data about the newest element in Hofstede's framework, i.e. "indulgence versus restraint".

The investigative approach of multi-level analysis is continued and elaborated by Taras, Kirkman and Steel (2010), who assess the impact of Hofstede's model (the first version) that consisted of four cultural dimensions (Hofstede, 1980) by collecting and retrieving data from 598 previously conducted studies, encompassing the expressed opinions of over 200,000 persons. In this way they meta-analyze the relationship between Hofstede's initial set of national culture dimensions and a deliberately designed variety of outcomes, describing important nuances in the organization's existence (performance, relations, attitudes, etc.). Their findings may be summarized as follows:

- The individual level of analysis is characterized by the similar strength with which values predict outcomes.
- Personality traits and demographics for certain outcomes as job performance, absenteeism, and turnover show significantly higher predictive power in comparison to the cultural values. The opposite situation is observed, concerning other outcomes as organizational commitment, identification, citizenship behavior, team-related attitudes, feedback seeking.
- Cultural values display different strength in their relationships with certain outcomes, arranged in a consecutive order by the observed decrease in this strength, i.e. emotions, attitudes, behaviors, and job performance.
- Stronger relationships between cultural values and outcomes are ascertained for managers, older, male, and more educated respondents.

- Their statistical analysis confirms significantly stronger effects in culturally tighter, rather than looser, countries.

Furthermore, Baskerville (2003) gives evidence of the striking pattern of citations from Social Sciences Citation Indices for the model of national cultural differences, provided by Hofstede (1980). The author identifies diverse application spheres of the framework and labels them as "disciplines". He states that Dutchman's findings show a continuous increase in citations in all disciplines, since these were first published and up to the moment of conducting his research which is not the traditional pattern of observed citations for the majority of studies, characterized by peaks of popularity about 3 to 5 years after publication, gradual decreases up to the tenth year after it and steady levels of citing from this time point on (Gamble, O'Doherty, & Hyman, 1987, p.18). The scientist reports great use of Hofstede's cultural dimensions in business-related research and psychological research and low use of it in anthropology and sociology (see table 5). In decreasing order by the number of attributed articles "management", "business administration" and "organizations" are the most popular sub-spheres.

Limiting the interested stakeholders of Hofstede's framework only to the academic constituency represents another fruitful approach in investigating the ways in which the cultural model is applied. This is accomplished by Sondergaard (1994) whose choice may be explained by the passing over of just a decade from Hofstede's first widespread publication where the respective questionnaire was presented publicly, so other scientists could test it and later on share their results,

*Table 5. Analysis of journal articles that cited Hofstede (1980) (and its later editions)*

| Topic | Total for topic area | 1981 | 1982 | 1983 | 1984 | 1985 | 1986 | 1987 | 1988 | 1989 | 1990 | 1991 | 1992 | 1993 | 1994 | 1995 | 1996 | 1997 | 1998 |
|---|---|---|---|---|---|---|---|---|---|---|---|---|---|---|---|---|---|---|---|
| Cross-cultural | 165 | | | 1 | 3 | 1 | 6 | 7 | 5 | 11 | 11 | 11 | 17 | 12 | 13 | 19 | 9 | 18 | 21 |
| Psychology | 540 | | 8 | 4 | 10 | 12 | 13 | 29 | 15 | 12 | 23 | 31 | 28 | 39 | 34 | 45 | 65 | 77 | 95 |
| *Business-related* | | | | | | | | | | | | | | | | | | | |
| Management | 238 | 2 | 3 | 9 | 6 | 8 | 12 | 10 | 4 | 2 | 10 | 15 | 9 | 14 | 22 | 29 | 26 | 17 | 40 |
| Business/Admin. | 232 | | 1 | 4 | 2 | 2 | 1 | 6 | 4 | 9 | 7 | 11 | 12 | 20 | 22 | 19 | 37 | 22 | 53 |
| Organizations | 126 | 2 | 4 | 2 | 2 | 1 | 4 | 1 | 5 | 5 | 10 | 5 | 4 | 6 | 13 | 10 | 12 | 16 | 24 |
| Marketing | 44 | 1 | | | | | | | 2 | | | | | | | 4 | 6 | 11 | 15 |
| Accounting | 31 | 1 | | | | | 2 | | 2 | 2 | | 2 | 3 | 1 | 3 | 4 | 6 | 11 | 15 |
| Operations Research, IT | 41 | 1 | | | 1 | | | | | | | 1 | | | | 5 | 8 | 9 | 16 |
| Sub-total for business-related | 712 | 6 | 9 | 15 | 11 | 11 | 19 | 17 | 17 | 18 | 27 | 34 | 30 | 44 | 60 | 71 | 92 | 79 | 152 |
| *Other than anthropology and sociology* | | | | | | | | | | | | | | | | | | | |
| Health/Medical | 30 | | | 1 | | 1 | | | 2 | 2 | 2 | | | 3 | 1 | 6 | 4 | 1 | 7 |
| Geography/Tourism | 17 | | | | | | | | | 2 | | | 1 | | | 2 | 5 | 2 | 5 |
| Law | 6 | | | | | | | 1 | | | | | | | | 2 | 1 | 0 | 2 |
| Economics | 17 | | | 1 | | | 1 | | | | | | | | 2 | 2 | 2 | 5 | 6 |
| Communication/Linguistics | 42 | | | | | 2 | 2 | 4 | 1 | 1 | | 3 | | 1 | 3 | 3 | 2 | 9 | 11 |
| Education/Child Development | 36 | | | | 1 | | 2 | 2 | 3 | 1 | 5 | 2 | | 1 | 4 | 2 | 4 | 1 | 8 |
| Various | 93 | 1 | 1 | 1 | 3 | 2 | 4 | 4 | 1 | | 5 | 3 | 8 | 5 | 11 | 11 | 6 | 10 | 17 |
| Sub-total for Other | 241 | 1 | 1 | 2 | 5 | 4 | 10 | 10 | 8 | 6 | 12 | 8 | 9 | 10 | 23 | 24 | 24 | 28 | 56 |
| Anthropology | 5 | | | | | | | 1 | | | | | | | | 1 | 0 | 0 | 3 |
| Sociology | 43 | 1 | | | 2 | 2 | 5 | | | 4 | 1 | 5 | 2 | 3 | 5 | 4 | 2 | 1 | 6 |
| Subtotal for anthropology and sociology | 48 | 1 | 0 | 0 | 2 | 2 | 5 | 0 | 1 | 4 | 1 | 5 | 2 | 3 | 5 | 5 | 2 | 1 | 9 |
| Totals | 1706 | 8 | 18 | 22 | 31 | 30 | 53 | 63 | 46 | 51 | 74 | 89 | 86 | 108 | 135 | 164 | 192 | 203 | 333 |

Source: (Baskerville, 2003).

experiences or propose recommendations for adaptation and/or improvement (see Hofstede, 1984). Sondergaard's mapping the use of the current cultural dimension set at this moment is revealed in table 6.

Soares, Farhangmehr and Shoham (2007) conduct a deeper investigation of Hofstede's dimensions use in international marketing studies. They confirm Lu et al. (1999) research results that there are three ways of applying the first five elements in this set, i.e. "to compare cultures, to support hypothesis, and as a theoretical framework for comparing

*Table 6. Hofstede's "Culture consequences..." (1984) and its applications for scientific use*

| Ways of applications | Main purpose of the users |
|---|---|
| 1. Nominal quotations | Mentioning a modern framework; 1036 quotations in Social Science Citations Index (SSCI) from 1983-1993. |
| 2. More substantively interesting citations | Covering remarks on Hofstede's ideas and results such as reviews and criticisms. |
| 3. Empirical usages | Making duplications or adjustment of Hofstede's framework by means of testing it with samples from different nations and/or regions and continuous tries to refine the associated constructs. |
| 4. Hofstede's concepts as a paradigm | Applying Hofstede's work without questioning its veracity without conducting a test or research on the respective ideas. |

Source: Sondergaard (1994).

cultures even if, in some cases, the actual scores are not used and the dimensions are measured with new or adopted instruments". In this way they reveal the relevance of these cultural dimensions for international marketing and consumer behavior, as follows:

- "Individualism – collectivism" influences innovativeness, service performance, and advertising appeals.
- Uncertainty avoidance impacts information exchange behavior, innovativeness, and advertising appeals.
- Power distance affects advertising appeals, information exchange behavior, innovativeness, and service performance.
- Masculinity impacts sex role portrays, innovation, and service performance.
- Long-term orientation influences innovativeness.

Without attracting the strongest scientific interest in itself (see Kirkman, Lowe, Gibson, 2006), there may be identified a research stream, oriented to uncertainty avoidance applications by authors in certain journals. For example, Rapp, Bernardi and Bosco (2011) adopt an interesting investigative approach to examine the use of Hofstede's uncertainty avoidance construct in international research among the scientific articles, published in the issues of "International Business Research" journal within a period of twenty five years, because the team of scientists posed the research question of how to determine the special features of use in which Hofstede's uncertainty avoidance construct has been incorporated into international research. In this way they created a sample of 118 articles and identified several research streams, differing by the specific use of this cultural dimension. These streams are arranged by diminishing number of the associated articles, as follows:

- The greatest number of articles (41) applied uncertainty avoidance dimension in order to explain formulated research hypotheses.
- The second group of articles (30) used this dimension as an independent or control variable.
- The third group of articles (29) utilized Hofstede's data to compare different countries through composite indices.
- The forth group of articles (15) applied this cultural dimension to support defended positions in the respective literature reviews of the scientific deliverables.
- The last group of articles (3) mentioned uncertainty avoidance in the research notes of their produce.

The aforementioned investigative point of view generates great value added in outlining not only important spheres of application for a certain component in Hofstede's national culture framework, but also it may serve as a milestone in design of further studies in the future.

Furthermore, almost the same team of researchers (Davis, Bernardi, Bosco, 2012) retained their attention to uncertainty avoidance cultural dimension, but this time investigated its application in the sphere of ethics studies and on the pages of Journal of Business Ethics, establishing a research period of 29 years. They found that the greater part of the reviewed articles used uncertainty avoidance strictly in the literature review (84%) and another smaller group of articles (16%) used it as a research variable or to compute a variable. Of course ethics research is not limited only to the aforementioned national culture dimension (see: Su, Kan, Yang, 2010; Tavakoli, Keenan, Crrijak-Karanovic, 2003; Tsui, Windsor, 2001).

The information technologies for sure constitute an interesting modern application sphere of Hofstede's model, outlining national culture dimensions, because: (a) *project view*: culture (at its different levels) has a great potential to influence the outcomes from design, implementation and use of information technology. (b) *process view*: since managerial processes are dependent on cultural factors, culture may directly, or indirectly, influence IT. For instance it turned out that:

- e-government's readiness and its components are related to dominating culture in certain countries and regions (Kovačić, 2005).
- Interesting differences are identified among multinationals, concerning manifestation of culture in the design of English-language and Chinese-language corporate websites (Chang, 2011).
- Based on Hofstede's cultural dimensions, it was found that microblogging seems to be more prevalent in emerging countries in comparison to developed ones. (Jobs, Gilfoil, 2012).
- National culture influences the patterns in software process mishaps that are revealed through incidents in global projects (MacGregor, Hsieh, Kruchten, 2005).

Even though Hofstede's framework is not the only one that is intended to serve as a means of surveying the impact of national culture on information systems domain, the first five of its components are widely used by the researchers which is very much in evidence by Ali and Brooks (2008), Myers and Tan (2003). Furthermore, Leidner and Kayworth (2006) reviewed and analyzed empirical and non-empirical IS-culture knowledgeable manuscripts, books and journals (for instance: MIS Quarterly, Journal of Management Information Systems, etc.) in order to classify them in six themes, as follows: (a) culture and information systems development; (b) culture, IT adoption, and diffusion; (c) culture, IT use, and outcomes; (d) culture, IT management, and strategy; (e) IT's influence on culture; and (f) IT culture. It appeared that the greater part of the scientific deliverables surveyed culture at national level and the majority of them utilized one or more of Hofstede's dimensions (see table 7).

Furthermore, Ford, Connelly and Meister (2003) make direct conclusions not only about dominating application spheres in information system (IS) research for Hofstede's framework, i.e. issues related to IS management and to IS, but also about the issue domains that at the moment of their survey seem relatively unexamined, i.e. IS development and operations and IS usage. They also find that theory development is not a prime objective for the scientists in the information system domain who used cultural dimensions model.

*Table 7. The intersection of national culture with the information system research*

| Citation, Methodology and Measure of National Culture | Relevant Finding(s) |
|---|---|
| **Information Systems Development** | |
| Hunter and Beck (2000) <br> Field study interviews (using Repertory Grid Analysis) of 70 Canadian and 17 Singaporean Respondents <br> Hofstede's Cultural Indices (1980): (PD), (UA), (IC), (MF). | Differences found across cultures in how excellent systems analysts are perceived. Excellent analysts from Singapore (high collectivism, low UA) are perceived to follow a more technocratic, dominant approach to clients while Canadian analysts (high individualistic, moderate-low UA) follow a more participative approach. |
| Keil, Tan, Wei, Saarinen, Tuunainen, and Wassenaar (2000) <br> Matching lab experiments in Finland, Singapore and Netherlands <br> Hofstede's Cultural Indices: (UA). | Cultures low in uncertainty avoidance (Singapore) exhibited greater tendencies to continue with troubled IT projects since their perceived risk was lower than with high uncertainty avoidance cultures. |
| Tan, Smith, and Keil (2003) <br> Matching lab experiment in Singapore and U.S. <br> Hofstede's cultural Indices: (IC). | Individualistic cultures amplify the impact of organizational climate on predisposition to report bad news (compared to collectivism) whereas collectivism strengthens the impact of information asymmetry on predisposition to report bad news (compared to individualism). |
| **Interorganizational Relationships** | |
| Steensma, Marino, Weaver, and Dickson (2000) <br> Five country survey of SMEs <br> Hofstede's Cultural Indices: (UA), (MF), (IC). | The tendency for SMEs to form technology alliances with others is greatest in countries that rate high in uncertainty avoidance and high in femininity (e.g., Mexico). SMEs in countries with collectivist values (Indonesia, Mexico) are more likely to form technology alliances involving equity ties than SMEs in more individualistic countries (Australia). |
| **IT Adoption and Diffusion** | |
| Garfield and Watson (1998) <br> Descriptive case study (content analysis) of government national information infrastructure (NII) archives across 7 countries <br> Hofstede's cultural Indices: (UA), (PD). | National culture plays a significant role in the development of a NII. Seven-country study suggests that countries will follow similar NII development models (family, village market, pyramid of people, or well-oiled machine) based upon similar cultural values related to uncertainty avoidance and power distance. |
| Griffith (1998) <br> Laboratory experiment comparing U.S. and Bulgarian student GSS teams (technology) <br> Hofstede's culture Indices: (PD). | Findings demonstrate that Bulgarian students (lower power distance) were more likely to report being dissatisfied with the GSS outcome than were the U.S. students (with higher power distance). |
| Jarvenpaa and Leidner (1998) <br> Single site case study (semi-structured interviews) of Mexican firm <br> Hofstede's culture Indices: (IC), (UA). | Mexican information services company succeeded despite presence of certain cultural barriers (e.g., high uncertainty avoidance and collectivism). Results show how managerial actions to shaped resource-based competencies led to shaping/recreating an information culture receptive to the information services industry. This transformation of culture led to greater levels of diffusion/acceptance of company's information services products. |
| Srite (2000) Field study of foreign students from 33 countries. <br> Hofstede's culture Indices: (UA, PD, IC, MF). | Individuals from high power distance countries were found to be less innovative and less trusting of technology. |

*Table 7. The intersection of national culture with the information system research (continued…)*

| IT Management and Strategy | |
| --- | --- |
| Burn, Saxema, Ma, and Cheung (1993)<br>Delphi study of 98 senior IT managers in Hong Kong<br>Hofstede's cultural indices: (UA), (IC), (PD), (MF). | Findings suggest that cultural values may influence the types of IS issues perceived to be most critical by IT managers. |
| Husted (2000)<br>Archival data analysis from Business Software Alliance (BSA)<br>Hofstede's culture indices: (IC), (UA), (PD), (MF). | Results indicate that software piracy is less prevalent in more individualistic (as compared to collectivist) cultural settings. |
| Kettinger, Lee, and Lee (1995)<br>Survey of IS users from Korea, Hong Kong, U.S., and Netherlands<br>Hofstede's culture Indices: (IC), (UA), (PD), (MF), (TO). | Study found that the service quality dimensions of the IS function differs across national cultures. Specifically, valid Information service quality (SERVQUAL) dimensions for Hong Kong and Korean were significantly different than for the U.S. and the Netherlands. |
| Shore, Venkatachalam, Solorzano, Burn, Hassan, and Janczewski (2001)<br>Survey of students from New Zealand, Hong, Kong, Pakistan, and U.S.<br>Hofstede's culture Indices: (IC), (UA), (PD), (MF). | Findings suggest that cross-cultural values influence attitudes toward intellectual property rights. Students from high power distance countries perceived less of an ethical issue with soft lifting (copying software for personal use). Students from high masculinity and individualistic cultures perceived more of an ethical problem with software piracy violations while those from high UA countries did not. |
| … | |
| **IT Use and Outcomes** | |
| Calhoun, Teng, and Cheon (2002)<br>Survey of Korean and U.S. professionals Cultural indices<br>by Hofstede (1980), Hofstede and Bond (1988), and Hall (1976) | High context culture respondents (Korea) experienced much higher levels of information overload from IT use on operational decisions as compared to respondents from a low context culture (U.S.). |
| Choe (2004)<br>Survey of Korean and Australian firms<br>Hofstede's culture Indices: (IC), (UA), (PD), (MF), (CD). | Under a high level of AMT, the positive effects of AMT and information (nonfinancial performance and advanced cost-control information) on the improvement of production performance is greater in Korean than in Australian firms. |
| Chung and Adams (1997)<br>Comparative survey of U.S. and Korean business firms<br>Hofstede's cultural Indices: (IC), (PD), (UA), (MF). | Comparison of respondents from significantly different national cultures (Korea and U.S.) resulted in no significant differences in group decision making behaviors attributable to Hofstede's four dimensions of culture. |
| Downing, Gallaugher, and Segars (2003)<br>Interpretive field study of Japanese and U.S. organizations<br>Hofstede's culture Indices: (IC), (UA), (PD), (MF). | Japanese companies (high uncertainty avoidance and collectivist) tend to select more information rich, socially present forms of media (face-to-face, fax, and phone) to facilitate empowerment whereas U.S. companies (low uncertainty avoidance and individualistic) tend to select more lean (efficient) forms of electronic media (e-mail, groupware, intranets) to facilitate empowerment. |
| Johns, Smith, and Strand (2003)<br>Survey of 78 MNCs Hofstede's culture Indices: (IC), (UA). | MNCs with lower uncertainty avoidance cultures are more likely to embrace new technologies and to encounter fewer impediments to international data flow. |
| … | |
| Source: (Leidner, Kayworth, 2006). | |

Table 8. Important comparisons among three cultural frameworks

**Journal and sample**

| Journal | Period available in ISI | Sample | % |
|---|---|---|---|
| Journal of International Business Studies | 1976 - 2011 | 1,176 | 32.3 |
| Management International Review | 1966 - 1990 2008 - 2010 | 891 | 24.5 |
| Journal of World Business | 1997 - 2011 | 394 | 10.8 |
| International Marketing Review | 1999 - 2010 | 315 | 8.7 |
| International Business Review | 2005 - 2011 | 231 | 6.3 |
| Journal of International Marketing | 1995 - 2011 | 319 | 8.8 |
| International Journal of Research in Marketing | 1997 - 2010 | 313 | 8.6 |
| **TOTAL** | | **3,639** | **100** |

Note: articles published in the period comprising the sample. % of total sample.
Source: Data collected from *ISI Web of Knowledge*. Computations by the authors.

**Ranking of the references of the cultural models**

| Journal | Hall | Hofstede | Trompenaars |
|---|---|---|---|
| Journal of International Business Studies | 897th (6) | 1st (213) | 94th (27) |
| Management International Review | 704th (2) | 5th (28) | 704th (2) |
| Journal of World Business | 228th (6) | 1st (76) | 11th (18) |
| International Marketing Review | 23rd (17) | 1st (62) | 61st (10) |
| International Business Review | 245th (5) | 2nd (52) | 91st (9) |
| Journal of International Marketing | 111th (8) | 2nd (59) | 181st (7) |
| International Journal of Research in Marketing | - (0) | 8th (21) | 430th (3) |

Note: In parentheses, the number of articles citing the work.
Source: Data collected using *ISI Web of Knowledge*, computations by the authors.

**Longitudinal analysis**

| | 1976–1984 | 1985–1993 | 1994–2002 | 2003–2011 |
|---|---|---|---|---|
| Hofstede (1980) | 5 | 33 | 150 | 323 |
| Trompenaars (1993) | - | 0 | 19 | 57 |
| Hall (1976) | 0 | 2 | 8 | 34 |
| **TOTAL** | **5** | **35** | **177** | **414** |

Source: Data collected from *ISI Web of Knowledge*.

Source: (Rosa dos Reis, Ferreira, Santos and Serra, 2013).

The great significance of Hofstede's framework may be better outlined if compared to other competing models in the cultural studies, struggling for the attention of business practitioners and academia. Rosa dos Reis, Ferreira, Santos and Serra (2013) stick to this purpose while conducting their bibliometric study of the cultural models, applied in the sphere of international business. In fact the team of researchers concentrates its attention to the three most popular models. i.e. Hofstede's cultural dimensions (1980), Edward Hall's high and low context culture (1976), and Trompenaars' seven dimensions of culture (1993). They surveyed the information of published articles from the seven most distinguished journals in the respective scientific field, available on ISI Web of Knowledge (see table 8). In this way they prove that Hofstede's set of cultural dimensions attracts the greatest numbers of citations, its popularity increases within the surveyed time period and many streams in the sphere of international business research are interlinked with it.

The presented analysis reveals the forming diversity in the use of Hofstede's framework and permits clear identifying and logic summarizing of its main streams of
application that emerged and are not considered as alternative ones. These streams are
depicted on figure 1.

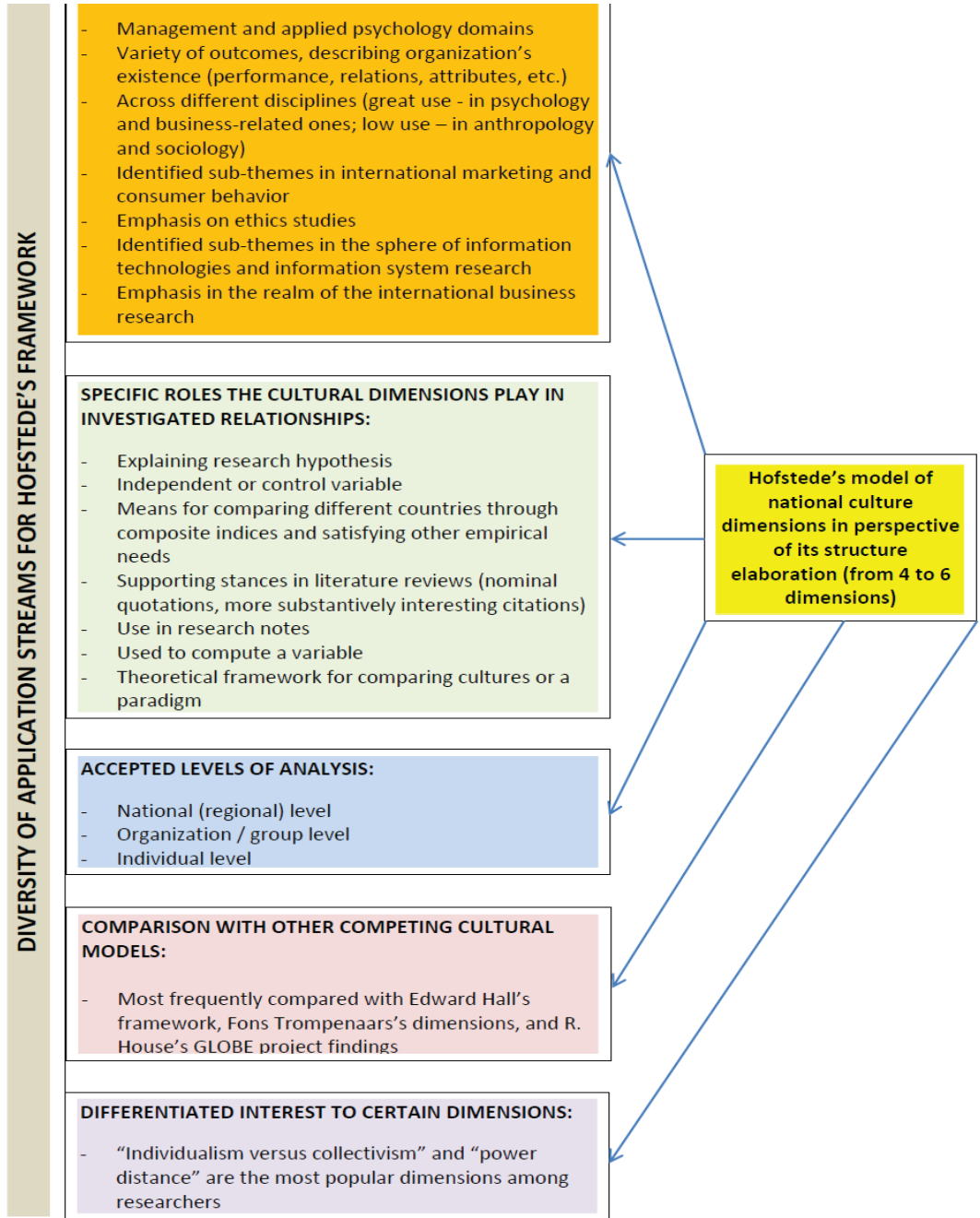

*Fig. 1. Mapping the main application streams for Hofstede's model*

## Main streams of criticisms to the national culture dimensions

The existence and the further elaboration of Hofstede's model are accompanied by unceasing flow of criticisms, oriented to the embedded assumptions in it by the team of contributors (see Hofstede, Hofstede, Minkov, 2010). Some of the critiques against the prominent scholar's work at least partially lost their relevance within framework's period of existence because of changes in Hofstede's doctrine (see Minkov, Hofstede, 2011) or are considered to some extent dissolved, absolved or resolved with passing over of time according to the subjective stances, occupied by the respective critics.

Brendan McSweeney (2002) is one of the most prominent and profound critics of Dutchman's findings who sets up his arguments along perceived "crucial methodological assumptions", incarnated in the analyzed model (see table 9).

*Table 9. Brendan McSweeney's appraisal of cultural dimensions set*

| Hofstede's assumptiopns… | Respective arguments against them… |
|---|---|
| The existence of three discrete and durable components (occupational, organizational and national cultures) | The existence of cultural heterogeneity in organizations. The survey is conducted among the employees of just one organization. Unclear definitions of applied constructs as 'practice' and 'perceptions of practice'. The sources/causes of the differences at the organizational level between practices or perceptions of practices are not addressed. Treating of cultural levels as methodologically distinct. Occupational cultures are not characterized by permanent programming. Social and institutional attributes are considered consequences of national culture. |
| The national is identifiable in the micro-local (all individuals in a nation or a 'central tendency') | Assuming national uniformity is not appropriate for a study that purports to have found it. The average tendency based on questionnaire responses from some employees in a single organization is not the national average tendency. |
| National culture creates questionnaire response | Differences identified on the basis of national stratification may not be treated as a consequence of national culture. Individual questionnaire respondents may not be accepted as relays of national culture. Survey's reliability is doubtful because IBM administered it and owned its results. |
| National culture can be identified by response difference analysis | Undisciplined mixing of two notions of culture - as a force, and as a decipherable manifestation. Inaccurate and incomprehensive descriptions of cultural manifestations of underlying national values. The composition and the number of the dimensions are questioned. There is no evidence for equivalence of meaning for dimensions across cultures. |
| It's the same in any circumstances within a nation | The apparent derivation of a national generalization from situational specific data is in fact a presupposition. The analyzed surveys encompassed only certain categories of IBM employees. The questions were oriented predominantly to workplace issues. The formal workplace was the only environment in which the survey was conducted. Generalizations about national level culture from an analysis of sub-national populations are not correct investigative approach. Validity of Hofstede's results is undermined by elusiveness of culture. In addition to national culture other types of cultures possess certain influence. Different levels and types of culture may interact. |
| Source: (McSweeney, 2002). | |

McSweeney's methodological perspective of critique may be enriched to some extent by an earlier analysis of Hofstede's findings, generated by Sondergaard (1994) who relying on other researcher's stances:

- Expresses his doubts in the validity of attitude-survey questionnaires as a means of providing inference about values.
- Gives a voice to his hesitations whether Hofstede's cultural dimensions may be considered as 'artifacts' of the analysis period.

Myers and Tan (2003) concentrate their critique of Hofstede's work on the concept of "national culture" by expressing their arguments against the appropriateness of "nation-state" as a unit of analysis and the possibility of concisely describing each country's culture with the help of a cultural dimension set. This stance allows them to define several key issues, related with the use of "national culture" construct, as follows:

- The relative newness of 'nation-state' phenomenon as a way of organization is pointed as a main reason of potential unreliability in Hofstede's model which is elucidated with historical evidence of the political development in the world during the nineteenth and twentieth century.
- The instability of form and makeup for the "nation-state" is emphasized as the second reason of potential unreliability in Hofstede's model which is supported by evidences from the recent history of two types: (a) these, associated with political unrest and sharp clashes (i.e. the collapse of the Soviet Union and Yugoslavia, and (b) these, associated with the potential effects of globalization and respective immigration flows to developed countries (i.e. displayed different extents of embracing certain cultural values and basic assumptions of the host culture by the newcomers).

- The researchers reject any obligatory alignment between a nation-state in its meaning of a political entity, and culture, providing examples of existing states without availability of any common basis in race, language, or culture (i.e. India, Switzerland, Yugoslavia, etc.).
- The contemporary anthropological view that is supported by the scientists, forces them to abandon Hofstede's static view of culture by defining it as an attribute that is "contested, temporal, emergent", "…interpreted, re-interpreted, produced and reproduced in social relations".
- The continuously accumulating research evidences, showing the extreme complexity and mediocre explanation by Hofstede's model of the relationship between "national" cultural values and culturally-influenced work-related values and attitudes.

Reviews of new streams in criticisms of international management emerged as a new source of posing arguments against Hofstede's cultural dimension set (Prasad, Pisani, Prasad, 2008). This is evident from the cited article by Ailon (2008). He applies an interesting research approach to deconstructing Hofstede's book "Culture consequences…" (the version with five cultural dimensions) within the framework of organizational discourse, i.e. analyzed in terms of its own proposed value dimensions. In this way the author reveals examples of how some non-Western societies seem devalued in this publication while some Western ones are idealized. Thus, the scientist explains the urgent needs to look for a solution to the problem of representing 'others', to appreciate political awareness in theory development in this stream of management and finally to reconsider important conceptualizations, dominating in related cross-cultural research.

Concentrating one's critiques only to a key dimension emerges as a new option of bringing arguments against Hofstede's framework. Tony Fang (2003) deliberately chooses as a target the fifth national culture dimension and in fact outlines six reasons of justifying his specific scientific stance:

- The researcher opposes Hofstede's unconsciously embedded association between: (a) 'short-term oriented' values and 'negative' values, and (b) 'long-term oriented' values and 'positive' values. Fang considers it as a tough violation of the Chinese Yin Yang principle (i.e. a philosophical flaw).
- He states that the respective meanings of the forty Chinese values set in the Chinese Value Survey (CVS) that constitute the fundamental of this dimension overlap with each other entirely or are highly interrelated. That is why the effect of bipolarity between values along this dimension may not be appraised (tested).
- The researcher detected unbalanced inclusion of values, stemming from basic religions in China (i.e. overreliance on Confucianism, but not on Taoism and Buddhism) in the values list, building the fifth dimension which is explained by Hofstede's choice to base his work on Chinese Culture Connection (1987).
- Detected linguistic issues, concerning some values, may have caused difficulties and inaccuracies in received results and their interpretations in conducted cross-cultural surveys.
- Students as a research object may not be associated with the average cultural values, possessed by typical members of respective national cultures.
- Different factor analysis technique is applied here and questionnaires are filled in by students instead of IBM employees.

Arbitrary accumulation of heaps from arguments by different scholars against Hofstede's cultural dimension set is another approach of criticizing the Dutchman (Ofori, Toor, 2009). In this way the authors provide a review of critiques selected by them for the purpose of enriching their deliverable's literature review or justify the need of conducting a certain research, intended to adapt or perfect Hofstede's framework in its application within a certain milieu by providing clear recommendations, certain steps, methods, etc. In this way they create a great bundle of miscellaneous criticisms:

- Observed overlapping in reflected values between the fifth dimension (Confucian dynamism) and individualism.
- The low percent of other scientists, interested in this framework who studied the fifth dimension, is explained by its inherent philosophical, language and methodological weaknesses (i.e. use of students, and use of different factor analysis techniques) that are ascribed to it.
- Detected sampling design issues.
- Accepting Hofstede's work as an attempt to measure the immeasurable (i.e. culture).
- Observed greater emphasis on proving one's own viewpoint rather than evaluating the adequacy of one's findings.
- Identified issues, related with model's operationalization, generalizability of the findings, author's subjectivity (i.e. culture-bound conclusions are made).
- The action research is not accomplished step-by-step.
- Left with the impression that a powerful feeling of ownership for the cross-cultural field is expressed by Hofstede et.al.
- Hofstede's model is viewed as an inhibitor for satisfying the need to look beyond it (i.e. identifying other values specific to certain regions, cultures, religions and countries).

This trend of analyzing Hofstede's model gains other supporters from the academic community. For example Froholdt and Knudsen (2007) produce a research on popular critiques and achieve similar results. But in this article only the critiques, enriching Ofori and Toor's (2009) "pile of arguments against Hofstede", are presented:

- The scientists provide us with a classification of the authors who express their doubts in Dutchman's findings: (a) radical criticizers who reject the framework; (b) constructive criticizers who are merely oriented to correction or refinement of the model; (c) relatively passive criticizers who warn against overuse of the aforementioned dimensions set that often is done in a simplified and uncritical way.
- A bunch of methodological problems are outlined, because: (a) questionnaires are used to measure self-representation, but not practices; (b) bipolarization is accepted as a main paradigm in the construction of these cultural dimensions; (c) other scientists reach other outcomes when deliberately 'unzipped' some of these dimensions.

The same approach was partially adopted by Jones (2007) who streamlines his critiques to Hofstede's four dimension model version through the perspective of traditional issues, associated with cross-cultural research as frequently arising semantic problems with used definitions, persistent adherence to methodological simplicity, and the bias to assume equivalency in phenomena occurrences in its functional, conceptual, instrument and measurement aspects. In addition to the abovementioned critique items as relevancy of used research instrument, the assumption of cultural homogeneity, the acceptance of national divisions, and overreliance on one company approach Jones (2007) incorporates new shades of meaning for some of them or formulates new ones, as follows:

- The results along dimensions as "masculinity - femininity" and "uncertainty avoidance" may be considered subject to dominating political influences at the time of the survey as deep memories of World War II, the on-going Cold War, and communist insurgence in Asia, Africa and Europe, because the constructed sample missed data from socialist counties and many Third World Countries.
- Considering the effects of driving forces as rapidly changing global environments, internationalization and convergence Jones does not miss the opportunity of expressing a widespread opinion by his colleagues that such survey does not create value added for the contemporary world.
- The scientist shows his hesitations whether cultural differences may be adequately explained by means of a model, consisting of four or five cultural dimensions.
- The use of the same questionnaire item on more than a single scale is considered not to be sufficiently supported by appropriate numbers of subjects (cases).

Furthermore, Catalin (2012) mentions several of the abovementioned critiques but brings forward the issue of Hofstede's accent on cultural differences and his lack of attention to cultural commonalities. The scientist outlines the fact that in Dutchman's model there may be found some coincidences in cultural dimension scores between an Eastern country and a Western one.

Preparing summaries of cited national culture values dimensions for the purpose of gaining deeper insights in the information system domain may be indicated as another source of special critiques, partially oriented to Hofstede's framework (Ali, Brooks, AlShawi, 2008). The arguments, aiming at the intersection between cultural studies and information system research, may be summarized as follows:

- Hofstede's complete attributing of detected differences between the respondents to national culture differences does not stimulate scientists to investigate specific influence(s) that different cultural levels (i.e. organizational, group, task force, etc.) may exert on studied information system phenomena.
- Dynamic nature of culture is neglected in Hofstede's model which may reduce the quality of surveys, investigating any potential influence of culture on the implementation and use of information systems.

Constructing a network of the reasons for observed low citations rate of Hofstede's "Culture consequences…" (1980) in the domains of sociology and anthropology allows Baskerville (2003) to formulate a bit indirectly his criticisms to national culture dimensions set, as follows:

- Evident adherence to anthropology and sociology as a realm, confirming the Dutchman's ideas and deliberate pursuing of methodological closeness with George Murdock's research (1962, 1963) that was very popular at that time. In fact Hofstede aims at conducting a cross-cultural study in the sphere of commerce and business research while accepting the nation as a unit of analysis. Such approach explains chosen direction of his prime interest in finding the ways in which national characteristics may be one variable in the analysis of business institutional or organizational behavior. But validity of this research strategy looks doubtful since human societies are not characterized by existence in isolation from each other and demonstration of only local variations. That is why according to Baskerville there is no guarantee that designed sample contains all known variants and barriers do not hamper meaningful comparisons between separate countries.

- The equation of nation states with cultures which traditionally is rejected in sociology and anthropology because: (a) percentage point differences cannot be always treated as evidence; (b) it is not acceptable all individuals inquired within a certain area to be grouped together under a dummy variable, labeled "country"; (c) frequently arising difficulties in making a difference between dependent and independent variables; (d) measured properties are not always characterized by stability. Furthermore, the "indices of culture" structure, proposed by Hofstede, does not imply any consideration of potential heterogeneity or suspected lack of independence of the unit of analysis.
- The quantification of culture is based on numeric dimensions and matrices. At the moment of Hofstede's research in the sociology and anthropology spheres there were no predecessors, utilizing indices, attributed with fixed numeric measures.
- Baskerville rejects Hofstede's stance that preferred status for the observer is not clearly defined in sociology and anthropology and presents evidence through the widespread use of fieldwork methods in these domains.
- The observed relationship of indices to other national data (i.e. social, political or economic measures). This fact allows Baskerville to express his opinion that these cultural dimensions "reflect mechanisms of social organization, or strengths and opportunism of different nations" that may originate from a nation's history.
- A single reverberation of personal confrontation against Hofstede becomes evident only when the criticizer forgets his good manners by insulting the Dutch scientist through a statement, inserted in the introduction of his article that read: "he (i.e. Hofstede) might not have studied culture at all" (Baskerville, 2003, p. 2).

Defending their scientific positions in return to Hofstede's critiques of GLOBE survey (2006), Javidan, House, Dorfman, Hanges and de Luque (2006) not only reveal and analyze the advantages of their research methodology, but also uncover an important insufficiency in Dutchman's work, i.e. his partial view to the relationship "national culture – national wealth". They consider that it is not enough to pose only the right question of 'What are the consequences of economic wealth?', but also state that each ambitious scientist is obliged to ask further 'What drives economic prosperity?' which is ignored by Hofstede.

The strong resonance from the scientific conflict between Hofstede's national culture dimensions and McSweeney's critiques to them (2002) is intelligently used by Williamson (2002) who occupies the role of an arbiter in this "ideas clash", delineating his moderate position by giving respective honor and unbiased appraisal of the respective contributions to both sides. In this way he seems to support the existence of several issues, related with the abovementioned model, labeled as "important warnings" for users:

- Not to assume culture uniformity for granted, i.e. all individuals from a given culture do not homogeneously possess the same cultural attributes.
- Not to assume cultural background as the only reason, explaining individuals' values or behavior.
- To distinguish between cultural constructs and their approximate measures, i.e. the scores of respective cultural dimensions.

Because of detected inherent contradictions, partial overlapping in meanings and simultaneous pursuing of many (sub-)directions among the components from the presented aggregate of critiques, directed to Hofstede's work, it may not be directly used as a means of clearly snap shooting the whole richness of outside views even after performing careful selection, analysis, logic classification and summary of these components. The tension of contradiction among different criticism items may be dissolved to some extent if Edgar Schein's (2004) concept of a group's fulfilling its ultimate need to balance between the interests of different constituencies for the sake of its own survival and successful development is transposed to Hofstede's cultural dimensions set, viewed as a product of a scientific team. In other words different criticizers may be determined as separate constituencies to the Dutchman's model which for sure has its own life, directed by Hofstede and associates' initial ideas and consecutive elaborations, and the continuous contributions of other researchers and consultants whose differing opinions and recommendations may be attributed to different milieus in which they applied the framework. Furthermore, the considerations of their criticisms by model's authors may be even interpreted as potential ways of continuously solving the problem of model's external adaptation to the current scientific and business environment. That is why it seems worth utilizing the fishbone diagram as an appropriate tool for analyzing the revolving the issue of "the arising, numerous critiques to Hofstede's dimensions" by looking for its potential sources and revealing the reasons of their occurrence (see figure 2).

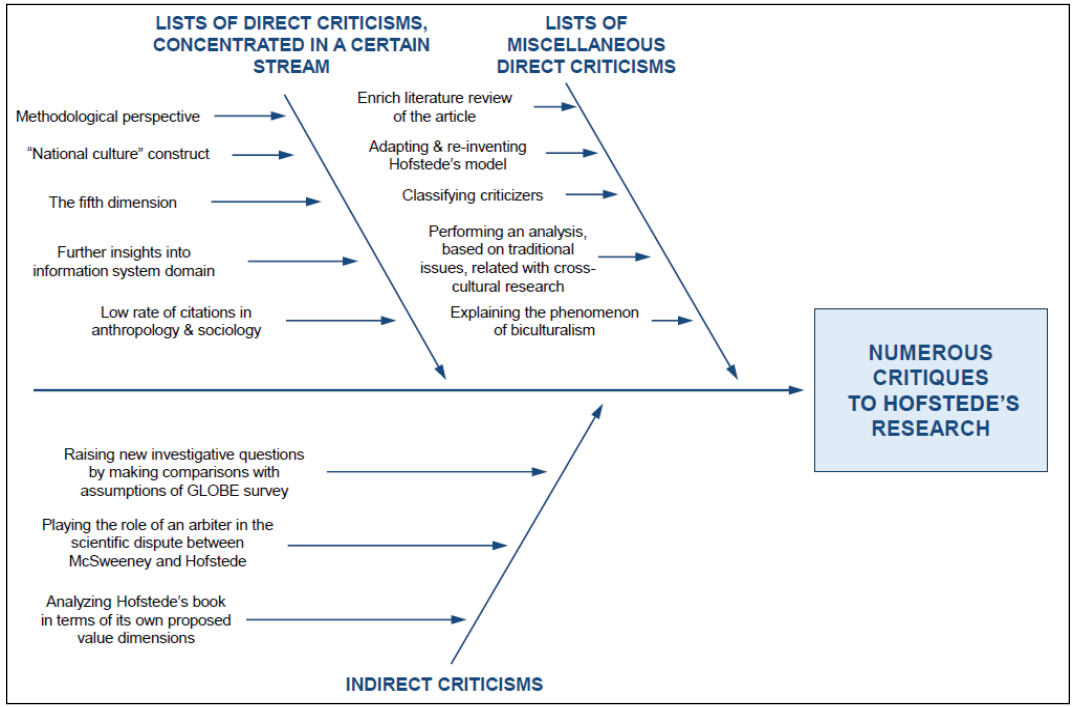

*Fig. 2. Mapping the causes of critiques for Hofstede's national culture dimensions*

## Conclusion

This contemporary snapshot of application spheres and most significant critiques for Hofstede's national culture model reveals the complexity of assumptions and paradigms, embedded in the initial construction and further elaboration of the dimensions set. The specific issues, encountered by different users, in the process of the larger penetration for this model in many scientific or business spheres as well as its deeper encroachment in some of them determine the large array of critiques the set has been attracting all these years. This situation allures me to analyze the model with the system approach that attaches great importance to these external forces as main drivers, pushing framework's further development and maintaining the interest of Hofstede et.al in it – an interest manifested by undertaking of key changes in the embraced doctrine. Such a mapping of the framework provides readers with: (a) a simple and clear explanation of its structure; (b) existing relations among its elements; (c) observed interactions with the higher-rank systems; (d) a useful means of making universalistic conclusions by beginners in the field, since most of the researchers are experts in boundary fields; (e) a generator of static pictures, revealing moment states of the model which dynamics may be traced by snap shooting successive photos (see figure 3).

Relying on clinical research in cultural studies, the author's position of an unbiased observer is considered appropriate in efficiently achieving the preliminary defined aims of this article. In this way an impartial view to model's being and becoming may be successfully obtained that is intended to be used by researchers, managers, consultants, and others to explain how things are done, thought, felt or perceived in relation to this framework by its diverse constituencies (see Schein, 2004). That is why the author's approach to reviewing and analyzing the set of national culture dimensions, its evolution, expressed through numerous elaborations – updating of meanings, adding new elements, applications at different cultural levels, incorporation in other models, and its acceptance levels and emerged application spheres – does not imply these options are mutually exclusive, but reveals them as an aggregate of different realms in which Hofstede's framework is needed to provide additional and plausible explanations of interesting societal and business-related phenomena. Thus the interested users in the set national culture dimensions may feel free to continue their creative use, elaborations and experiments with it, providing incessant pipeline of potential change proposals to the authors of the model who are granted the right to accept or reject them and/or follow their own scientific endeavors.

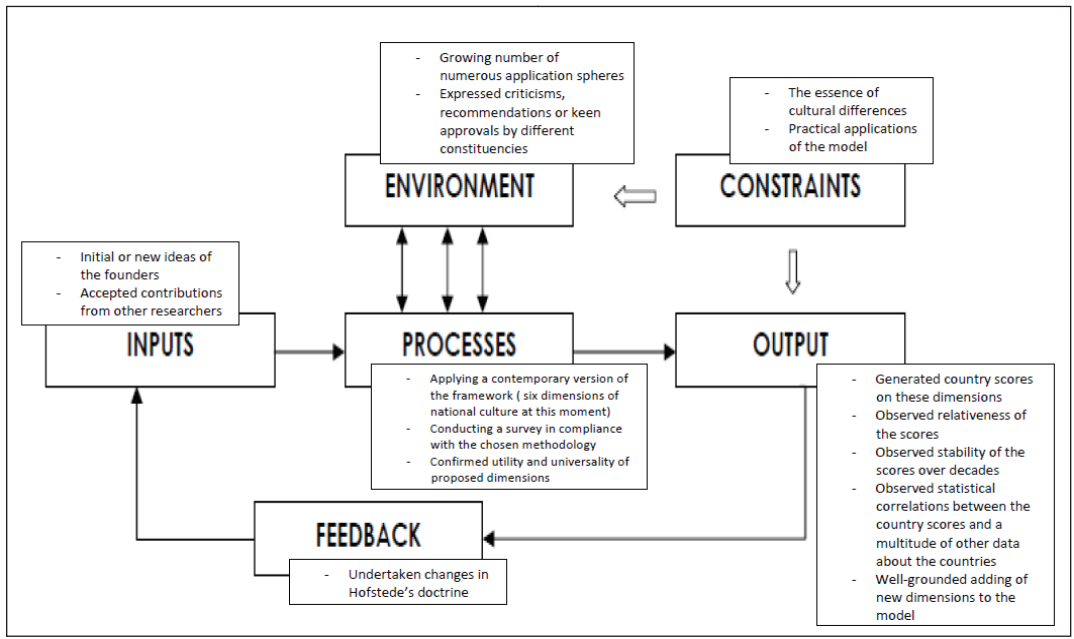

*Fig. 3. Hofstede's model of national culture dimensions as a system*

On one side, such an "intensive testing" may be regarded as a prerequisite for model's lasting life. On the other side, the readers are incited not be in a hurry to express their opinion in relation to this model by undertaking intrepid and complacent survey duplications in different regions or communities or due criticizing or giving recommendations to its structure, appropriate survey design, implementation process and application spheres, but first read with patience Hofstede's publications and after that deliberately explore how their intended research design and potential results may contribute to the current stage in development of this model, because even recent surveys apply its elder (outdated) versions, for example a set of four or five cultural dimensions.